\documentstyle[sprocl,epsfig]{article}
\def\be{\begin{equation}}
\def\ee{\end{equation}}
\def\bea{\begin{eqnarray}}
\def\eea{\end{eqnarray}}

\begin{document}
\title{Enhanced Charmless Yield in $B$ Decays and Inclusive $B$-Decay Puzzles}
\author{Isard Dunietz}
\address{\it Fermi National Accelerator Laboratory, P.O. Box 500,  
Batavia, IL 60510}
\maketitle\abstracts{
Our analysis suggests that the charmless yield in $B$ decays is enhanced over
traditional estimates.  The $c\overline c$
pair produced in $b\rightarrow c\overline cs$ transitions may be seen significantly as
light hadrons due to non-perturbative effects. Existing data samples at $\Upsilon
(4S)$ and $Z^0$ factories allow key measurements which are outlined.}

\section{Motivation}

One prime motivation for optimizing our understanding of inclusive $B$ decays is
CP violation.  CP asymmetries at the 50\% level are predicted for the
time-evolved $B_d \rightarrow J/\psi K_S$ decays~\cite{bigisanda}, within the CKM
model. 
The few hundred reconstructed $J/\psi K_S$ events~\cite{cdf} would thus
allow meaningful CP studies, once they are tagged. Tagging denotes
distinction of an initially pure $B_d$ and $\overline B_d$. An optimal tagging
algorithm combines self-tagging \cite{gnr,leptag,cdf} with all available information
from the other $b$-hadron decay~\cite{distinguish}. Thus inclusive $b$-hadron
decays must be understood. Such an understanding would enhance CP studies with
$B$ samples both inclusive \cite{bbdcp} or exclusive. It would reduce backgrounds
for any $B$-decay under study. Intriguing hadronization effects may be discovered~\cite{close}.

\section{Traditional Puzzles}

The $b$ is known to decay normally to a $c$, and that charm flavor is referred to
as ``right" charm.  In contrast, the $b\rightarrow \overline c$ process produces
``wrong" charm. The penguin amplitudes give rise to $b\rightarrow s$ transitions,
which are seen as a kaon, additional light hadrons, and possibly additional
$K\overline K$ pairs. Due to the small $|V_{ub}/V_{cb}| \sim 0.1$, the $b\rightarrow
u$ transitions are negligible at the present level of accuracy.
Theory calculates the rates for $b\rightarrow c\ell\overline\nu$ \cite{nir},
$b\rightarrow c\overline cs$ \cite{rud,bagan,voloshin}, and the ratio of rates
\cite{rud}
\begin{equation}
r_{ud} \equiv \frac{\Gamma (b\rightarrow c\overline ud')}{\Gamma (b\rightarrow
ce\overline\nu )} = 4.0 \pm 0.4\;.
\end{equation} 
The CKM parameters cancel in the ratio. The phase-space factor cancels in leading
order and $r_{ud}$ would be 3 because of color counting. QCD corrections (complete to next-to-leading-order with finite charm quark masss) have
been found to enhance this ratio to 4.0 \cite{rud}.
Of course we are not dealing with freely decaying $b$-quarks, but with decays of
$b$-hadrons. It must thus be emphasized that the calculation of $r_{ud}$ assumes local quark-hadron duality.

\hspace*{-3.5pt}
In this talk, $b$ denotes the weighted average of produced $\overline B$
mesons. The semileptonic BR is
\begin{equation}
BR_{s\ell} \equiv \Gamma (b\rightarrow Xe^- \overline\nu )/\Gamma (b\rightarrow
{\rm all}) ,
\end{equation}
and the charm multiplicity $\stackrel{(-)}{c}$ per $b$ decay is given by
\begin{equation}
n_c = \frac{\# \stackrel{(-)}{c}}{\#b} = 1-B(b\rightarrow {\rm no}\;{\rm charm})
+ B(b\rightarrow c\overline cs') \;.
\end{equation}
The current theoretical status is summarized in Fig.~1 \cite{neubert,neuberts},
which plots the theoretically allowed $(n_c, BR_{s\ell}$) region.

\begin{figure}
 \centering
 \mbox{\psfig{figure=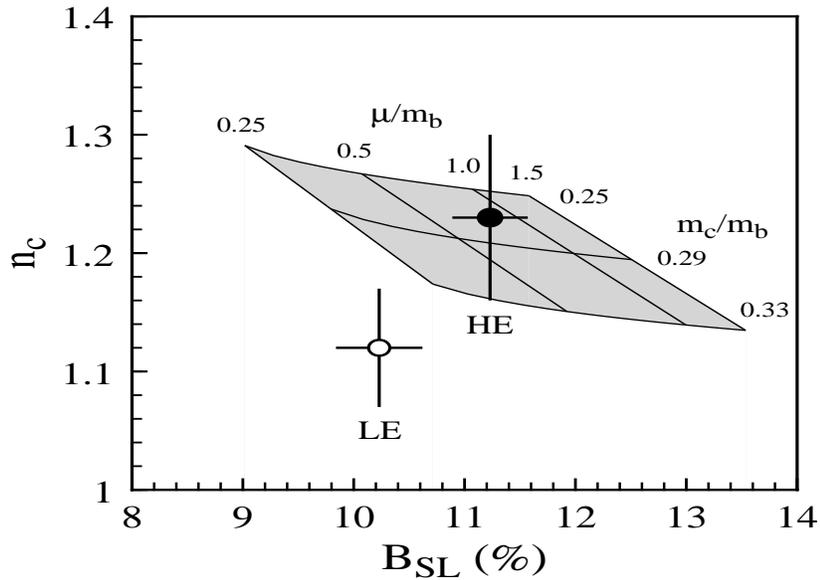,height=3.1in,width=4.5in}}
 \caption{Theoretical prediction for the semileptonic branching ratio and charm multiplicity. The data points show the average experimental values obtained at $\protect\Upsilon(4S)$ (LE) and $Z^0$ (HE) factories.  Figure taken from Ref.~\protect\cite{neubert}.}
 \label{brslnc}
\end{figure}

The low (high) horizontal curve is for a large (small) $m_c /m_b$ ratio. The
diagonal curves are given for various renormalization scales. The left boundary
is given by $\mu /m_b =0.25$, for which $r_{ud} \;\raisebox{-.4ex}{\rlap{$\sim$}} \raisebox{.4ex}{$>$}\;5$, see Figure 2.

The measured charm multiplicity per $B$ decay $n_c$ (as summarized in Fig.~1) must
be revised downward significantly, because of several reasons.  First, the measured central value of $\Xi_c$
production is too large. An upper-limit has been derived and is drastically
smaller~\cite{recal}. The drastic reduction can be traced back to a large enhancement in the 
absolute BR scale of $\Xi_c$ decays, a conclusion supported by recent 
work of Voloshin~\cite{VoloXic}. Second, the world-average for
\begin{equation}
B(\Lambda_c \rightarrow pK^- \pi^+)= 0.044 \pm 0.006
\end{equation}
must be sizably revised upward to $0.08 \pm 0.02$~\cite{recal,brlmbc}.
This causes $n_c$ to decrease more significantly at $Z^0$-factories (because of $\Lambda_b$ production) than at $\Upsilon(4S)$ factories.

\begin{figure}
 \centering
 \mbox{\psfig{figure=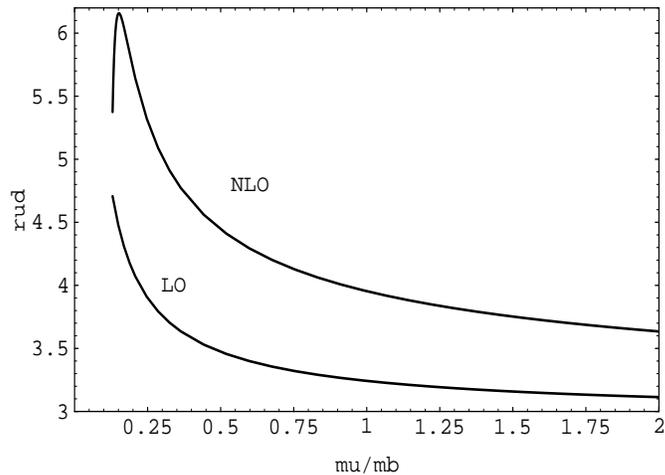,height=3.1in,width=4.5in}}
 \caption{Scale dependence of $r_{ud}$ for leading-order (LO) 
 and next-to-leading-order (NLO) approximations~\protect\cite{rud}.  
Figure taken from Ref.~\protect\cite{disy}.}
 \label{rudvsmu}
\end{figure}

However, $n_c$ and $BR_{s\ell}$ are not the only observables.  With the recent
flavor specific measurement of wrong-sign $\overline D \;[\overline D^0$ or $D^-
]$ production in $b$ decay, $B(b\rightarrow \overline D)$, the quantity $r_{ud}$ can now be experimentally
extracted,
\begin{equation}
  r_{ud} [{\rm exp.}]  = 
   \frac{  B(b\rightarrow {\rm open}\;\; c) 
         - B(b\rightarrow {\rm open}\;\;\overline c) 
         + B(b\rightarrow u\overline c s')}{BR_{s\ell}} -  2 - r_\tau \;,
\end{equation}
with minimal theoretical input, including\cite{disy,rtau}
\begin{equation}
   B(b\rightarrow u\overline cs')=0.0035 \pm 0.0018,\; 
   {\rm and}
\end{equation} 
\begin{equation} 
r_\tau =0.22\pm 0.02\;.
\end{equation}  
Using CLEO data alone $r_{ud}[{\rm exp.}] = 4.1 \pm 0.7$~\cite{disy}.

The sizable $b \to \overline D$ observation unearthed an overlooked background $b \to \overline D \to \ell^-$ in model-independent, inclusive $BR_{s\ell}$ measurements\cite{recal}.  The $Z^0$ measurement will be reduced significantly, and is more affected than the $\Upsilon(4S)$ measurements because of differences in cuts on the signal lepton momentum.  The model-independent extraction of  $BR_{s\ell}$ requires the removal of $B^0 - \overline B^0$ mixing effects and the value of the average mixing parameter $\overline\chi$ as input.
But both the value of $\overline\chi$ and the removal of $B^0 - \overline B^0$ mixing effects will have to be modified, because the secondary leptons $b \to \stackrel{(-)}{c} \to \ell$ experience different mixing than the primary leptons $b \to \ell^-$~\cite{recal}. We anticipate~\cite{recal} that reanalyses of data will significantly reduce the difference between the $BR_{s\ell}$ measurements from the $Z^0$ and  $\Upsilon(4S)$ environments in favor of the lower $\Upsilon(4S)$ result~\cite{cleosl}.

After applying the revisions onto Fig.~1, the experimental measurements from $\Upsilon (4S)$ and $Z^0$ factories are consistent.  The $\Upsilon (4S)$ data support a low renormalization scale $\mu$, and are marginally consistent with theory based on the heavy quark expansion~\cite{neuberts,neubert}.

\section{Flavor-Specific Input}

CLEO\cite{moriond} and ALEPH\cite{barate} determined
\begin{equation} B(b\rightarrow \overline D) = \Biggl\{\begin{array}{ll} 0.085 \pm
0.025 \;\;\;\;{\rm CLEO}\;\;\;{\rm 1996} \\ 0.145\pm 0.037 \;\;\;\; {\rm
ALEPH}\;\;\;{\rm 1996}\end{array} 
\end{equation}   
Do those measurements confirm
the prediction \cite{bdy} of $B(b\rightarrow \overline D) \sim 0.2$?

To answer that question, a synthesis of all available data, flavor-specific and
flavor-blind, was in order. The $B(b\rightarrow$ no open charm) is that fraction
of $\overline B$ decays which has no weakly decaying charm, that is, no separate
charm vertex. It can be inferred indirectly \cite{disy}:

{\flushleft {Method A:}}
\begin{equation} 
B(b\rightarrow {\rm no}\;{\rm open} \;{\rm charm}) =
1-B(b\rightarrow {\rm open} \; c)\;-B(b\rightarrow u \overline c s').
\end{equation}
{\flushleft {Method B:}}
\begin{equation} B(b\rightarrow  {\rm no}\;{\rm open} \;{\rm charm}) = R -
B(b\rightarrow {\rm open} \;\overline c)\;. 
\end{equation} 
Here, $R$ is the remaining
BR after reliable components have been subtracted, 
\begin{eqnarray} 
  R & \equiv &
  B(b\rightarrow {\rm no}\;{\rm charm})\;
  + B(b\rightarrow c\overline cs')\;+ B(b\rightarrow u\overline cs') = \nonumber \\ &
  = & 1-B(b\rightarrow c\ell\overline\nu ) 
  -B(b\rightarrow c\overline ud') =\nonumber \\ & = &
  1-BR_{s\ell} [2+ r_\tau +r_{ud}]\;. 
\end{eqnarray}  
Theory provides $r_\tau$ \cite{rtau}, 
$r_{ud}$ \cite{rud}, experiment $BR_{s\ell} =0.105 \pm 0.005
$\cite{cleosl}, and $R=0.35\pm 0.05$ results.                               
This result changes only minimally to
\begin{equation}
\label{r}
R=0.36\pm 0.05 ,
\end{equation}
once differences in the $B^-$ and $\overline B_d$ rates governed by $b\rightarrow
c\overline ud$ have been conservatively incorporated~\cite{neuberts,bbd}. Our
prediction Eq.~(\ref{r}) for $R$ combines the most accurate information available
from both theory and experiment~\cite{disy}.

The average of methods A and B is denoted by Method C:
\begin{equation}
B(b\rightarrow {\rm no}\;{\rm open}\;{\rm charm})\;=\frac{1}{2} [1+R-Y_{{\rm
open}\;c}\;-B(b\rightarrow u \overline c s')],
\end{equation}
where the flavor-blind quantity
$Y_{{\rm open}\;c} \equiv B(b\rightarrow {\rm open}\;c) +B(b\rightarrow {\rm
open}\; \overline c) \;.$
Because flavor-blind yields are better known than flavor-specific ones, Method C
allows the most accurate prediction for $B(b\rightarrow {\rm no}\;{\rm open}\;{\rm
charm})$. Note that while Method A involves experimental data alone (with
minimal theoretical input), Methods B and C require the theoretical prediction
for $r_{ud}$. Method C reduces its sensitivity on theoretical input with regard
to Method B, because of the factor $1/2$.
Table I summarizes our findings~\cite{disy}.

\begin{table}
  \caption{Indirect estimates of no open 
     charm in $B$ decays~\protect\cite{disy}}
  \begin{center}
  \begin{tabular}{|c|c|}
\hline
    Method & $B(b\rightarrow {\rm no}\;{\rm open}\;{\rm charm})$ [CLEO] \\
    \hline
    Method A & 0.15 $\pm$ 0.05 \\
    Method B & 0.17 $\pm$ 0.06 \\
    Method C & 0.16 $\pm$ 0.04 \\
\hline
  \end{tabular}
  \end{center}
\end{table}

Why is $B(b\rightarrow {\rm no}\;{\rm open}\;{\rm
charm})$ enhanced over traditional expectations of $0.05 \pm 0.01$~\cite{disy}. New physics may provide a solution and could enhance the charmless $b\rightarrow
s'$ transitions \cite{newphysics}.
But before concluding that, all Standard Model explanations must be exhausted
first.

Non-perturbative effects could be responsible for $c\overline c$ pairs to be seen
significantly as light hadrons. The $c\overline c$ pairs produced in $b\rightarrow
c\overline cs$ transitions have low invariant mass and are dominantly in a
color-octet state \cite{palmerstech,disy}.
The predominantly $c\overline c$ color-octet configuration may have sizable overlap
with the wavefunction of $c\overline c$- hybrids, $H_c$, which are made of $c\overline c$ and glue~\cite{kuti,isgurpaton,hybrid,Hc,bcs}.
Although their masses could be beyond the open charm threshold~\cite{kuti,Hc,bcs}, model-dependent selection rules
suppress $H_c \rightarrow D^{(*)} \overline D^{(*)}$ transitions~\cite{isgurpaton,selrul}. Consequently,
they could be narrow and could be seen sizably as light hadrons.  That light hadron yield is probably governed significantly by resonant production of light gluonic hadrons~\cite{close}. More generally, because the $b$-quark is sufficiently massive and decays in a gluon rich environment (provided, for instance, by the soft gluons emanating from the light spectator quark[s]), we anticipate copious production of gluonic hadrons~\cite{close} and enhanced non-perturbative annihilation of $c \overline c$ pairs (see Figure (1b) in Ref.~\cite{bjorken}).

\begin{figure}[t]
 \centering
 \mbox{\psfig{figure=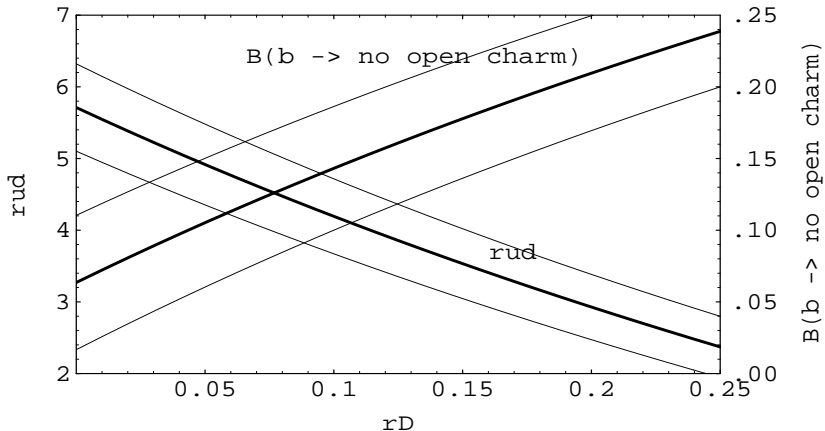,height=2.5in,width=4.5in}}
 \caption{$B(b \to$ no open charm) and $r_{ud}$ as 
functions of $r_D$~\protect\cite{disy}.}
 \label{nocvsrd}
\end{figure}

Perhaps the wavefunctions of light hadrons $[\pi ,\rho,
K^{(*)},$ etc.] have a non-negligible component of intrinsic $c\overline c$
\cite{brodsky,zhitnitsky}. The generic charmless mode is $\overline B\rightarrow \overline K n\pi\; 
(n \; \geq \; 1),$ where no
partial subset of final state particles reconstructs a charmed hadron. The $c\overline
c$ component may have transformed itself into an intrinsic piece of decay
products, and interference effects may be important~\cite{brodskythank}. Because more
excited light resonances have generally a larger intrinsic charm component than less excited states~\cite{brodskythank}, it
appears plausible that the $\overline B\rightarrow \overline Kn\pi$ processes feed
through such more excited resonances.\footnote{We expect those resonances to have net zero
strangeness, else the whole invariant mass $m_b$ of the $b\rightarrow c\overline cs$
process would be available to create strange resonances with intrinsic
charm.} The end result of such a scenario is very similar to the above mentioned possibility of
charmed hybrid production. Nevertheless, they could be distinguished.

Charmed hybrids are predicted~\cite{kuti,Hc,bcs} to have masses of about 4 GeV or above, while light resonances
with an intrinsic $c\overline c$ component could be significantly lighter. 
Consequently, a detailed momentum spectrum of the recoiling $K^{(*)}$ in such $B$
decays may help in differentiating the various possibilities. A surplus of very high momentum $K^{(*)}$ is consistent with the production of excited
resonances that contain intrinsic charm or with direct production of light gluonic hadrons. A high momentum $K^{(*)}$
excess (although less high than the aforementioned) may indicate $H_c$
production, while the momentum spectra of produced kaons in non-resonant
$\overline B\rightarrow \overline K n\pi$ processes will be different.
Such and other
non-perturbative effects must be carefully investigated.

Another solution is provided by a reduction of $B(D^0\rightarrow K^-\pi^+)$ from presently accepted values,
which would
increase $n_c$ and would cause $B(b\rightarrow {\rm no}\;{\rm open} \;{\rm charm})$ to
decrease towards traditional expectations \cite{recal,distw}. This and other
systematic effects have been discussed in Ref.~\cite{disy}.

Figure 3 emphasizes
the importance of accurate measurements of
\begin{equation}
r_D \equiv\frac{B(b\rightarrow \overline D)}{B(b\rightarrow D)}\;.
\end{equation}
That figure plots $B(b\rightarrow {\rm no}\;{\rm open} \;{\rm charm})$ (Method A)
and $r_{ud}$ as a function of $r_D$ using essentially only experimental input.

The ALEPH measurement fully reconstructs both charm mesons in $\overline
B\rightarrow D\overline DX$ transitions, and thus suffers from low statistics~\cite{barate}.
The existing data samples at $Z^0$-factories allow more accurate $B(b\rightarrow
\overline D)$ measurements. After selecting an enriched $b$-sample, one needs to
reconstruct a \underline{single} $\stackrel{(-)}{D}$ only, employ optimized
flavor-tagging, and correct for $B^0-\overline B^0$ mixing effects.
(We add parenthetically that those data samples allow meaningful CP violating
tests~\cite{bbdcp}.) If sizable charged $B^\pm$ data samples can be efficiently
isolated, one could determine again $B(B^- \rightarrow \overline DX)$ and $B(b\rightarrow \overline D)$ without
the need for a flavor-tag and for corrections due to $B^0-\overline B^0$ mixing.
The accurate determinations of $B(b\rightarrow \overline D)$ are crucial for
resolving the inclusive $B$ decay puzzles (see Figure 3), and should be pursued with high
priority.

\section{Conclusions}

Under the traditional assumption of a tiny $B(b \to$ charmless), the accurately measured $BR_{s\ell} =0.105\pm 0.005$ \cite{cleosl} allowed the
prediction \cite{bdy}
\begin{equation}
n_c =1.30\pm 0.05\;,
\end{equation}
while experimentally \cite{ncnew}
\begin{equation}
n_c=1.10\pm 0.05\;.
\end{equation}
Recent flavor-specific measurements opened up new aspects pertaining to this
puzzle and allowed the indirect extraction of $B(b\rightarrow {\rm no}\;{\rm
open}\;{\rm charm})$ in a variety of ways. The results of the methods are
consistent, strengthening our conclusion that the charmless yield in $B$ decays is enhanced over traditional estimates. 
Method C yields the most accurate prediction of
\begin{equation}
B(b\rightarrow {\rm no}\;{\rm
open}\;{\rm charm}) = 0.16\pm 0.04 \;.
\end{equation}
This large charmless yield would show up as an enhanced fraction of $b$-decays, without
a separate daughter charm vertex.
We expect the underlying physics to be non-perturbative in nature, which causes a sizable fraction of $c\overline c$ pairs to be
seen as light hadrons.  The momentum spectrum of the involved $K^{(*)}$ may help in distinguishing among the various scenarios. 

We touched upon the systematics of our analysis and considered the
parameters $[B(b\rightarrow {\rm no}\;{\rm
open}\;{\rm charm}), r_{ud}, B(D^0 \rightarrow K^-\pi^+ ), r_D]$ and correlations
among them~\cite{disy}. The prediction for $r_{ud}$ involve larger theoretical
uncertainties than presently realized~\cite{disy}. [Under the assumption of local duality, the dependence of the predicted $r_{ud}$ on the scale $\mu$ is large, and is not improved by going from leading-order to next-to-leading-order, see Figure 2.  While the large scale dependence is troublesome, an even more disturbing aspect is the fact that duality assumes an inclusive rate based on 3 body phase-space, while the
$b\rightarrow c\overline ud$ transitions proceed sizably as quasi-two body
modes.\footnote{The $b\rightarrow c\overline ud$ transitions could be modelled as
follows. For small invariant $\overline ud$ masses $(m_{\overline ud}\;\leq \;m_\tau )$, the color-singlet $\overline ud$
pair hadronizes with little or no final state interactions. The factorization assumption can
be justified, because by the time the $\overline ud$ forms a sizable color dipole [with which it could interact with its surrounding environment], it left the other debris of the $B$-decay far behind~\cite{colortransparency}.
 The hadronization of those $\overline ud$ pairs can be determined from the
well-studied $\tau$ decays, $\tau \rightarrow \nu +\overline ud$, which are dominated by the production of $\overline u d$ resonances. The $b\rightarrow c$
transitions can be modelled by HQET with input from semileptonic measurements and are seen dominantly as $(D, D^*, D^{**}$) resonances.
Factorization is not as reliable for higher invariant $\overline ud$ masses.
Fortunately, the $\overline ud$ invariant mass spectrum falls rapidly off at higher masses, as shown by a straightforward Dalitz plot. Assuming
factorization, the vector contribution can be inferred from $e^+e^-$ measurements
at the same c.m. energy, where the isospin 1 component has to be isolated from the data. The
axial-vector component can be obtained from the relevant spectral function. We
are in the process of developing a $b\rightarrow c\overline ud$ Monte Carlo simulation~\cite{dru}.}]  Fortunately, $r_{ud}$ can be
extracted from experimental measurements alone, which can be confronted with
theory.
More accurate determinations of $r_D$ or equivalently $B(b\rightarrow \overline
D)$ are possible from existing data samples at LEP/SLD/CLEO.  They are invaluable in
guiding us toward a more complete understanding of $B$-decays.

$B$ decays are a fertile ground for searching and discovering subtle 
had\-ronization effects.  By utilizing the long lifetime of $b$-hadrons,  vertex detectors can drastically reduce backgrounds.  To fully explore multibody decays of $b$-hadrons it will be essential to not only have good $\pi/K/p$ separation, but the ability to detect $\pi^0, \eta^{(')}, \gamma$ as well. An additional very important bonus will be a more optimal exploration of sizable CP violating effects residing in such multi-body $B$ decay modes.  Especially striking effects within the CKM model are expected in $b \to d$ transitions.

\section*{Acknowledgments}
This work was supported in part by the Department of Energy, Contract No.
DE-AC02-76CH03000.

\end{document}